# Bringing Engineering into the Physics Lab: Exploring Crank-Slider Dynamics with Smartphones

*Josep Ll. Suñer[1], Rod Milbrandt[2], Juan C. Castro-Palacio[3], Alfonso Merchante[1], Francisco M. Muñoz-Peréz[3], Juan A. Monsoriu[3,*]*

[1]Instituto de Ingeniería Mecánica y Biomecánica, Universitat Politècnica de València, 46022 València, Spain.

[2]Faculty of Physics and Engineering, Rochester Community and Technical College, Rochester, Minnesota, USA

[3]Centro de Tecnologías Físicas, Universitat Politècnica de València, Camino de Vera, s/n, 46022 València, Spain

*Corresponding author: *jmonsori@fis.upv.es*

Many of the students enrolled in the calculus-based physics sequence are pursuing engineering degrees. Lab activities that have a direct connection to engineering, analyzed with the physics they have been studying, can grab students' interest and provide a real-world connection. All the better if the equipment is inexpensive and partly self-built.[1] Allowing students to focus on the underlying physics while exploring systems similar to those encountered in engineering practice can further increase the educational value of such activities.

In addition, we wanted to use smartphones as data collection devices. Smartphones have become valuable tools for physics education because they provide an accessible and inexpensive means of experimental data acquisition through their built-in sensors. [2] Several free applications, such as *Physics Toolbox*[3] and *phyphox*[4], facilitate sensor-based data collection and analysis, making it possible to perform meaningful laboratory activities without expensive equipment. Numerous studies have demonstrated the educational value of smartphone-based experiments in mechanics, including investigations based on accelerometers and gyroscopes. [5,6] Students are generally familiar with smartphones and often enjoy using them as scientific instruments, which can further increase engagement with laboratory activities.

We chose to use a simple crank-slider mechanism for our activity. The crank-slider converts rotary motion into linear motion and has common applications in such things as internal combustion engines and compressors. Educational studies have shown that slider-crank mechanisms and related mechanical devices provide effective contexts for introducing engineering concepts and connecting theoretical models with practical applications. [7-9]

In our implementation, a rotary motor drives a disk connected by a rod to a cart moving along an air track. Students investigate the resulting motion using smartphone data acquisition and video analysis. Experimental measurements are then compared with the predictions of a mathematical model derived from the geometry of the system. The activity therefore combines smartphone data acquisition, video tracking, mathematical modeling, and engineering applications within a single laboratory experience, providing students with an opportunity to connect introductory physics concepts with a real engineering mechanism.

## Equipment

The experimental setup is shown in Figure 1. The disk used in our experiment has a diameter of 50 cm and is mounted on a rotary motor. The disk and support used in our implementation were 3D printed, although other materials such as wood could also be used. The simplicity of the design makes the apparatus easy to reproduce in instructional laboratories and allows students to focus on the physical principles governing the motion rather than on the construction details.

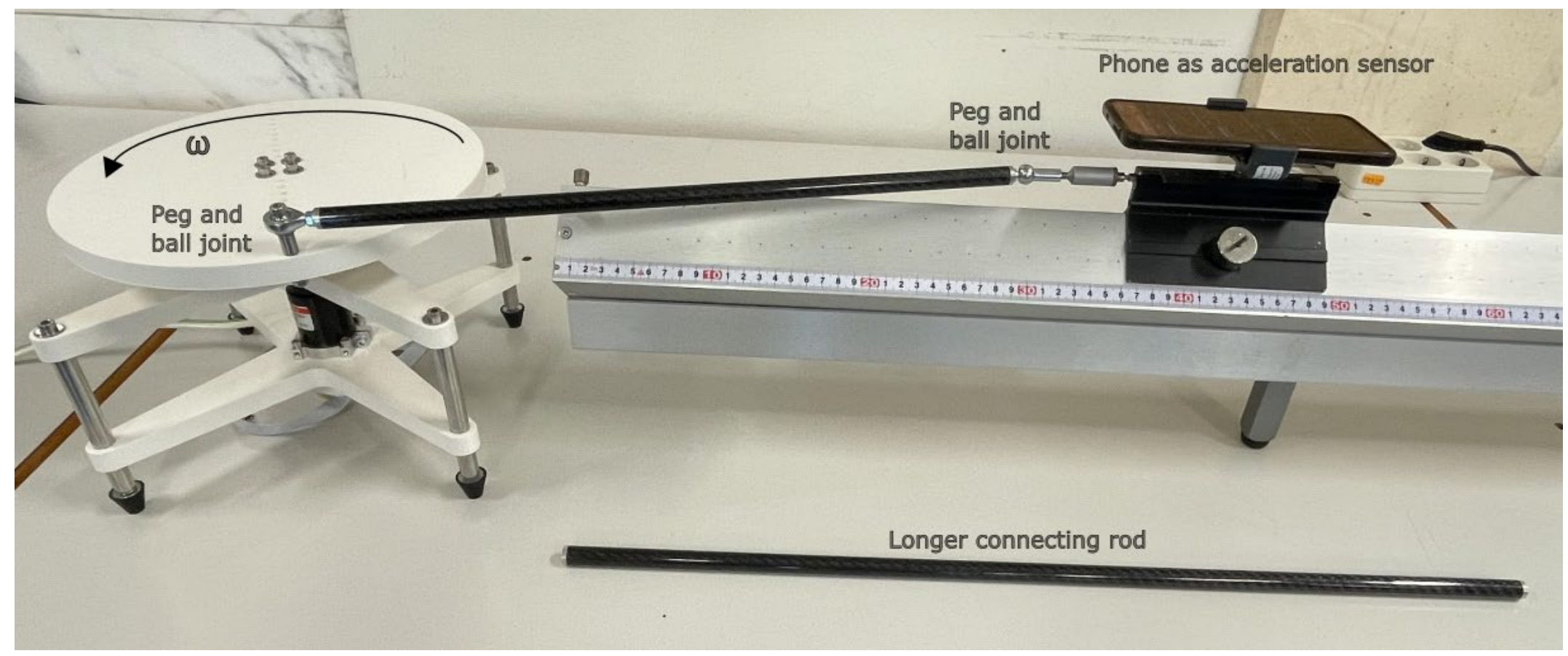

**Figure 1.** Experimental setup used to investigate the dynamics of a crank-slider mechanism. The main components are identified by the labels shown in the figure.

A row of holes was drilled along a diameter of the disk, allowing the effective crank radius to be adjusted in increments of 2 cm. This feature enables students to investigate how the geometry of the crank-slider mechanism influences the resulting motion. A peg inserted into one of these holes is connected through ball joints to a rod, which in turn is attached to an air track cart. The rotational speed of the motor is controlled by an adjustable DC power supply, making it possible to explore different operating conditions and observe their effect on the measured kinematic quantities.

Two smartphones are used during the experiment. One is mounted on the cart, as shown in Figure 1, and records acceleration data using a free application such as Physics Toolbox or phyphox. The second smartphone, positioned on a tripod, records a video of the cart motion from the side or from above. The use of readily available mobile devices eliminates the need for specialized data-acquisition hardware while providing measurements of sufficient quality for quantitative analysis.

For an experimental run, the air track is switched on, acceleration and video acquisition are started, and power is supplied to the motor. After several rotations, the motor is stopped and both the acceleration data and video file are exported to a computer for analysis. A typical acceleration record obtained with phyphox is shown in Figure 2.

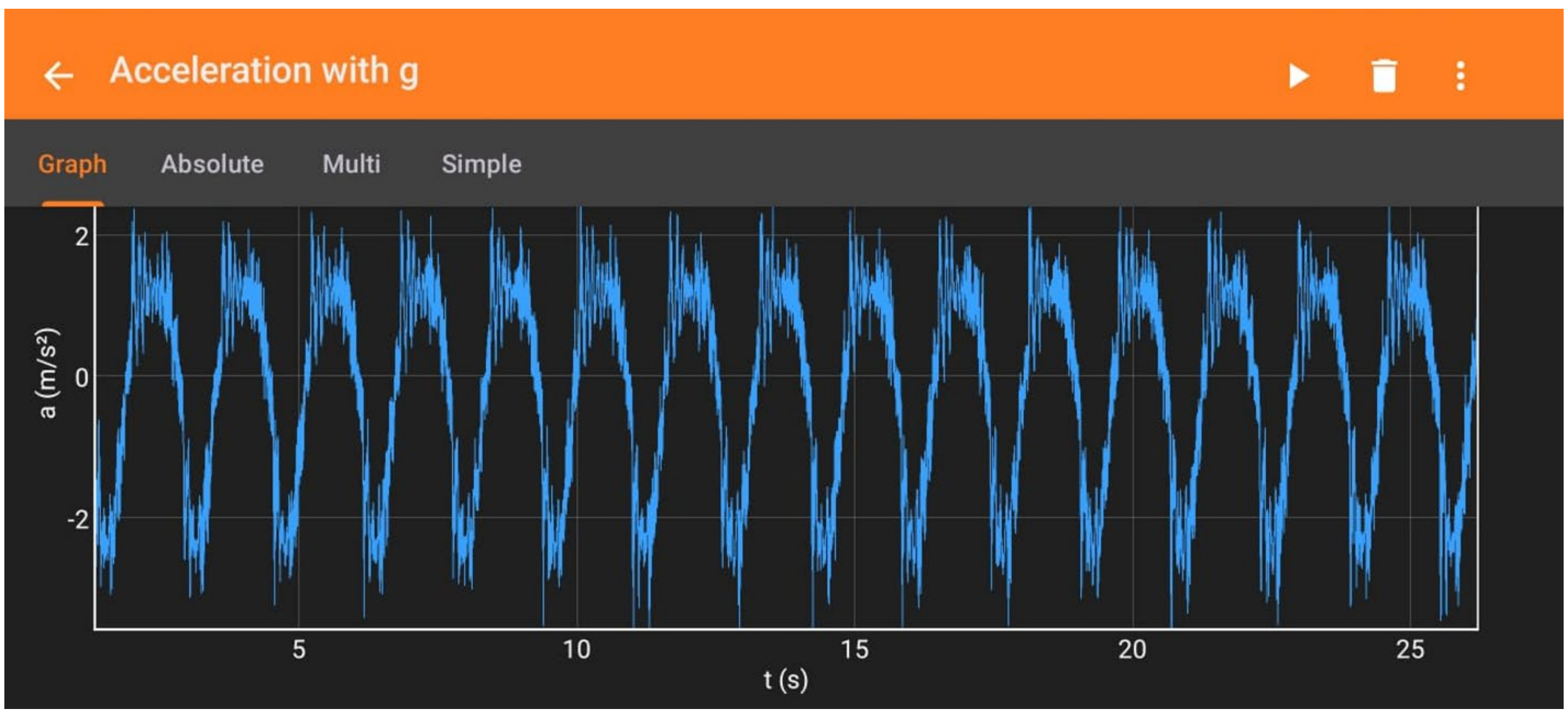

**Figure 2.** Typical acceleration-versus-time data recorded by the smartphone accelerometer using the phyphox application during a crank-slider experiment.

The video recorded with the second smartphone is analyzed using the free Tracker software, [10] and the resulting cart position data are exported to Excel for further processing. From these position measurements, students determine the corresponding cart mean velocities and mean accelerations by applying the finite-difference relations $\bar{v} = \Delta x/\Delta t$ and $\bar{a} = \Delta v/\Delta t$, respectively. This procedure provides a straightforward introduction to numerical differentiation and allows students to obtain kinematic quantities directly from experimentally measured positions.

A typical Tracker analysis is shown in Figure 3. The screenshot illustrates the video-tracking procedure used to obtain the experimental position data of the cart as a function of time. Numerical differentiation of these measurements yields acceleration data that are subsequently used for comparison with the theoretical model presented in the following sections.

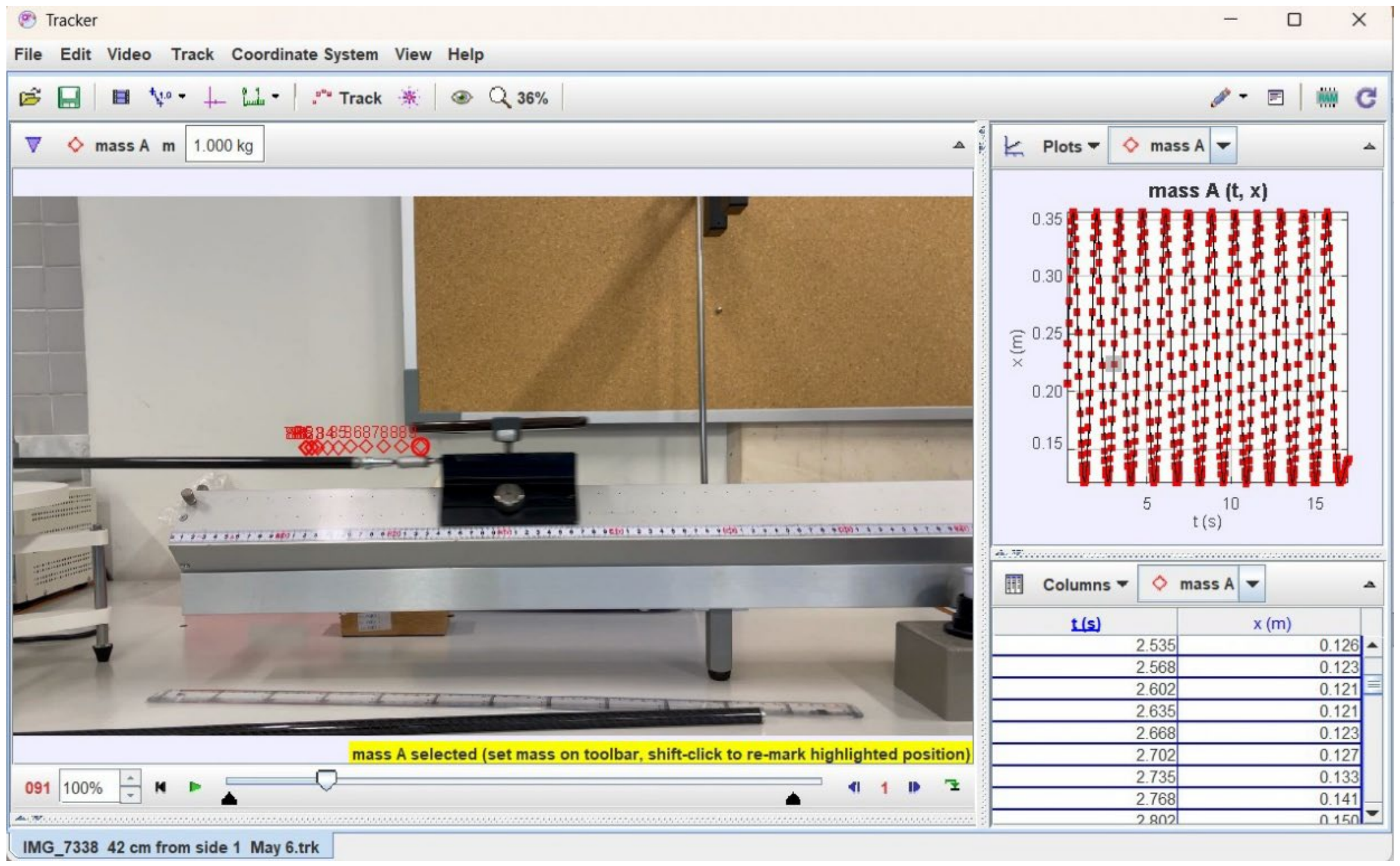

**Figure 3.** Typical Tracker analysis of the crank-slider motion. The software provides the cart position as a function of time, which is subsequently used to determine the acceleration through numerical differentiation.

## Theoretical Model

The geometry of the crank-slider mechanism is shown in Figure 4. The system consists of a crank of radius $R$ rotating at a constant angular velocity $\omega$ and connected to a cart through a rod of length $L$. As the crank rotates, the cart is constrained to move along a straight line, transforming the rotational motion of the motor into linear motion. This type of mechanism is widely used in engineering applications such as internal combustion engines, compressors, and pumps.

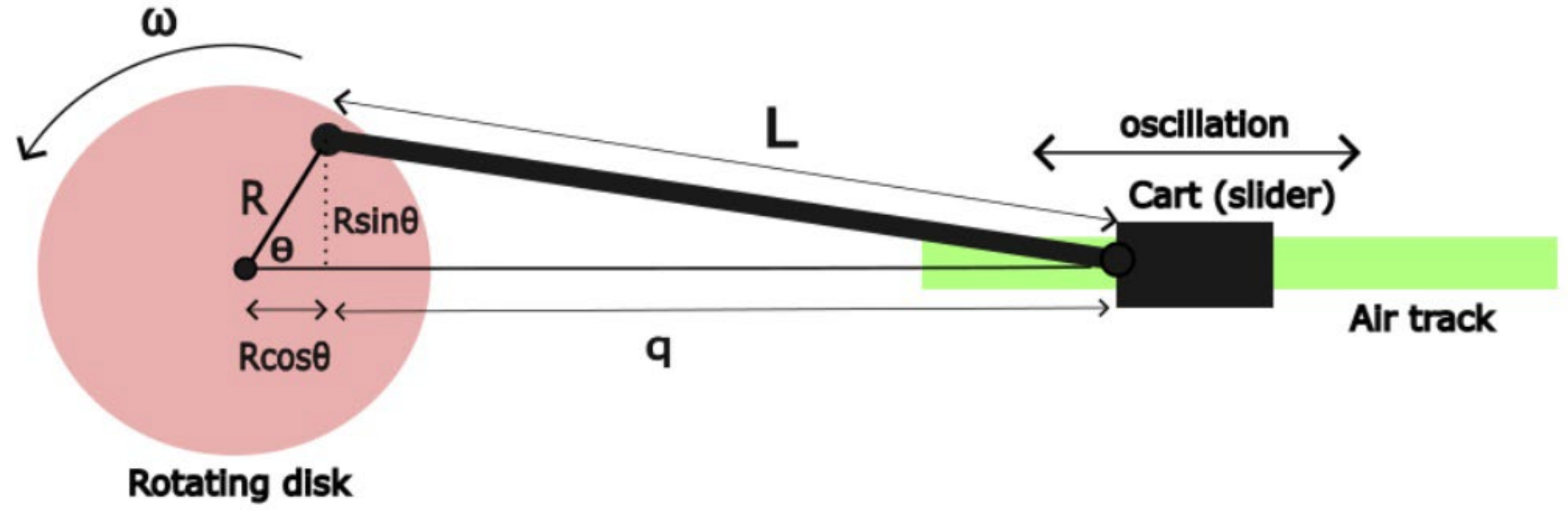

**Figure 4.** Geometrical model of the crank-slider mechanism used to describe the motion of the cart.

To analyze the motion of the cart, we first determine its position $x$. Referring to Figure 4, the cart position can be written as $x = q + Rcos\theta$, where $q$ represents the horizontal projection of the connecting rod and $\theta$ is the angular position of the crank. Therefore, the cart position is determined by the combined contributions of the crank and the connecting rod. Finding an expression for $q$ allows us to describe the motion of the entire system.

Using standard right-angle trigonometry, $q$ can be expressed as

$$q = \sqrt{L^2 - R^2 sin^2\theta}, \tag{1}$$

which follows directly from the right triangle formed by the connecting rod, its horizontal projection, and the vertical displacement of the crank pin. Substituting Eq. (1) into the position expression gives

$$x = Rcos\theta + \sqrt{L^2 - R^2 sin^2\theta}. \tag{2}$$

Equation (2) relates the cart position to the instantaneous angular position of the crank. Because the connecting rod has a finite length, this relationship is nonlinear and differs from the simple cosine dependence that would characterize ideal simple harmonic motion.

Since the disk rotates at a constant angular velocity, the angular position of the crank varies linearly with time according to

$$\theta(t) = \omega t + \phi, \tag{3}$$

where $\phi$ is a phase angle introduced to account for the initial position of the crank at the start of data acquisition.

Substituting Eq. (3) into Eq. (2) yields the cart position as a function of time,

$$x(t) = Rcos(\omega t + \phi) + \sqrt{L^2 - R^2 sin^2(\omega t + \phi)}. \quad (4)$$

Equation (4) completely describes the displacement of the cart. Although the crank rotates uniformly, the motion of the cart is not exactly sinusoidal. The finite length of the connecting rod introduces a geometric asymmetry that causes the cart to spend slightly different amounts of time in different regions of its trajectory. This departure from simple harmonic motion becomes more noticeable as the ratio $R/L$ increases

The cart velocity is obtained by differentiating Eq. (4) with respect to time

$$v(t) = -\omega R sin(\omega t + \phi) - \frac{\omega\, R^2 sin(\omega t + \phi)\cos(\omega t + \phi)}{\sqrt{L^2 - R^2 sin^2(\omega t + \phi)}} \quad (5)$$

The above equation predicts how the cart speed changes throughout the cycle. Unlike the velocity of a simple harmonic oscillator, the velocity profile is influenced by the geometry of the crank-slider mechanism and is therefore not perfectly symmetric. A second differentiation gives the cart acceleration,

$$a(t) = -\omega^2 R cos(\omega t + \phi) + \frac{\omega^2 R^2 \left(L^2 sin^2(\omega t + \phi) - L^2 cos^2(\omega t + \phi) - R^2 sin^4(\omega t + \phi)\right)}{\left(L^2 - R^2 sin^2(\omega t + \phi)\right)^{\frac{3}{2}}} \quad (6)$$

Equation (6) constitutes the theoretical model used to describe the acceleration of the air-track cart. The mathematical treatment requires only trigonometry and differentiation and is therefore accessible to students enrolled in a calculus-based introductory physics course.

One particularly interesting aspect of this model is that it links a realistic engineering mechanism with concepts typically introduced in introductory mechanics. Students can therefore appreciate how mathematical models arise from geometric considerations and how experimental measurements can be used to test theoretical predictions.

## Experimental Results and Model Validation

For the data run presented here, the crank radius was 12 cm and the connecting rod length was 42 cm. These values were fixed throughout the experiment. The theoretical acceleration model given by Eq. (6) was fitted to the experimental data using the Solver tool in Excel. The fitting parameters were the angular velocity $\omega$ and the phase angle $\phi$, while the geometric parameters $R$ and $L$ were fixed at their measured values.

The acceleration obtained from video analysis is presented first in Figure 5 together with the theoretical prediction. Although some fluctuations are present due to the numerical differentiation of the position data, the model reproduces the main features of the measured acceleration curve. In particular, it correctly captures both the amplitude and the non-sinusoidal shape of the oscillation predicted by the crank-slider geometry.

The fit shown in Figure 5 yielded an $R^2$ value of 0.95, indicating excellent agreement between the theoretical model and the experimental data. The resulting parameters were $\omega = 4.54\ rad/s$ and $\phi = 0.527\ rad$. These results demonstrate that the mathematical model derived from the geometry of the mechanism provides an accurate description of the cart motion.

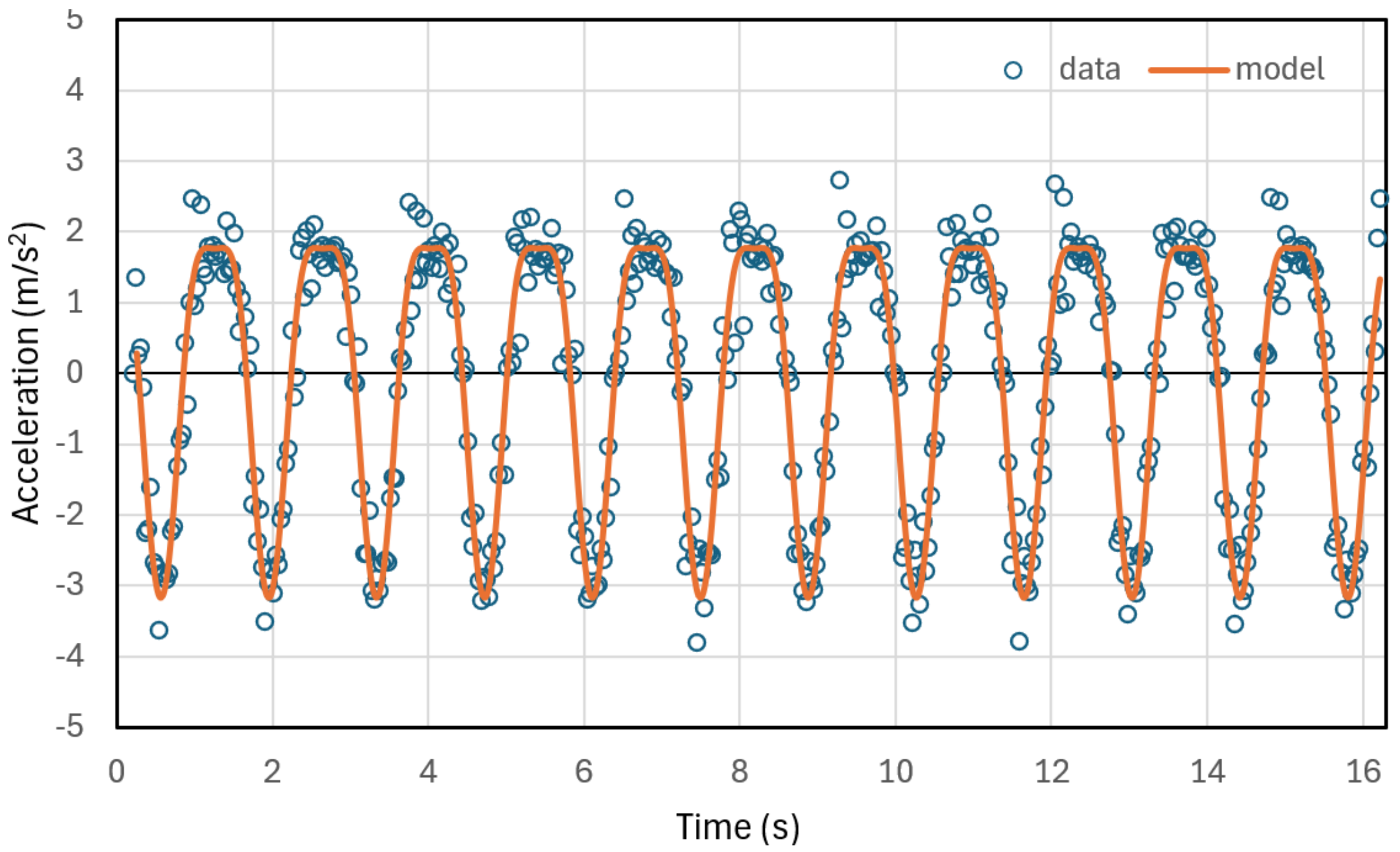

**Figure 5.** Comparison between the acceleration obtained from video analysis and the theoretical model for a crank radius of 12 cm and a connecting rod length of 42 cm.

The acceleration measured directly by the smartphone sensor was also analyzed using the same fitting procedure. The resulting comparison between experimental data and theoretical prediction is shown in Figure 6. Compared with the video-analysis results, the accelerometer data exhibit a higher level of noise, which is expected because the sensor records instantaneous accelerations and is more sensitive to vibrations and small mechanical imperfections in the system.

Despite this additional noise, the theoretical model again reproduces the overall behavior of the measured acceleration as seen in Figure 6. The fitted angular velocity was $\omega = 4.57\ rad/s$.

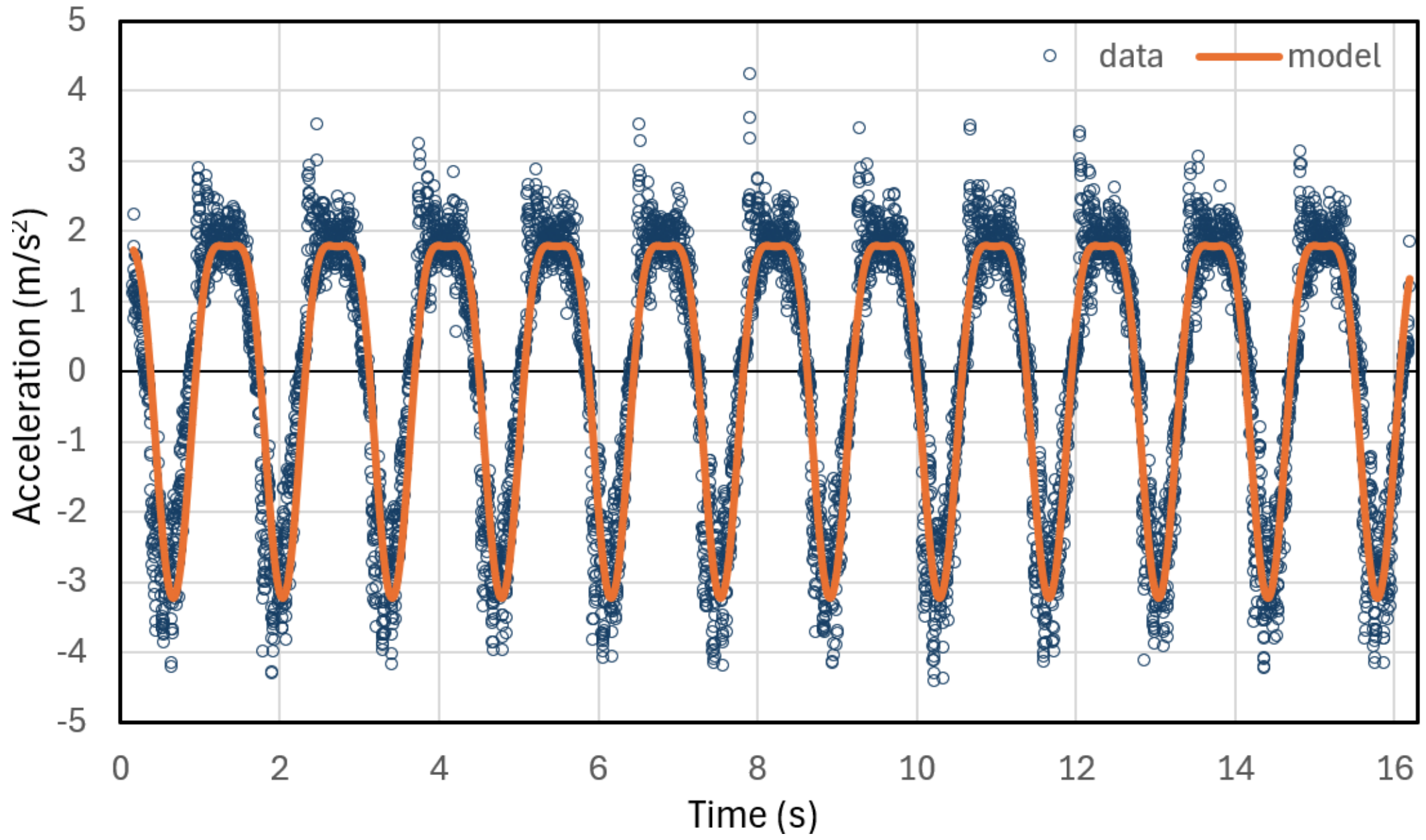

**Figure 6.** Comparison between the acceleration measured by the smartphone accelerometer and the theoretical model for a crank radius of 12 cm and a connecting rod length of 42 cm. $R^2 = 0.90$

The angular velocity obtained from the smartphone data differs by less than 1% from the value determined using video analysis. This close agreement between two completely independent measurement techniques provides strong evidence for the validity of both the theoretical model and the experimental methodology. The agreement obtained with both measurement techniques demonstrates that smartphones can provide reliable quantitative data for studying engineering-related mechanical systems in introductory physics laboratories. From an educational perspective, the comparison also gives students an opportunity to evaluate different experimental approaches, assess the influence of measurement noise, and appreciate the role of mathematical modeling in interpreting real-world data.

## Conclusions

The crank-slider mechanism is a common engineering system that can be studied using simple and inexpensive laboratory equipment. In this activity, students combine smartphone-based data acquisition, video analysis, and mathematical modeling to investigate the motion of a cart driven by a rotating crank.

The experimental results obtained from both video analysis and smartphone measurements showed good agreement with the theoretical model derived from the geometry of the mechanism. Furthermore, the close agreement between the two independent measurement methods provides students with an opportunity to evaluate experimental uncertainty and compare different approaches to data acquisition.

The activity can be easily modified by varying parameters such as the crank radius, connecting rod length, or rotation speed, allowing students to explore how these factors affect the resulting motion. In addition to reinforcing concepts from introductory mechanics, the experiment introduces students to mathematical modeling in a context directly connected to engineering applications.

## Acknowledgements

This work was funded by the Polytechnic University of Valencia (grants PIME/23-24/374 and PIME/25-26/578), Spain. The authors also thank the Institute of Educational Sciences at the Polytechnic University of Valencia (Spain) for its support of the EICE SmartSTEM Teaching Innovation Group, and Rochester Community and Technical College (MN) for sabbatical funding.